\documentclass[11pt,preprint]{aastex}

\slugcomment{}

\shorttitle{3-5 $\mu$m spectra of ULIRGs} 
\shortauthors{E. Sani et al.}

\begin{document}
\title{3-5 $\mu$m spectroscopy of obscured AGNs in ULIRGs \thanks{Based on observations collected at the European Southern Observatory,
Chile (proposals ESO 73.B-0574, 79.B-0052).}} 

\author{E. Sani,\altaffilmark{1} G. Risaliti,\altaffilmark{2,3} M. Salvati,\altaffilmark{2} R. Maiolino,\altaffilmark{4} A. Marconi,\altaffilmark{1} S. Berta,\altaffilmark{5}
V. Braito,\altaffilmark{6,7} R. Della Ceca\altaffilmark{8} and A. Franceschini,\altaffilmark{5} }

\email{sani@arcetri.astro.it}

\altaffiltext{1}{Dipartimento di Astronomia, Universit\`a di Firenze, Largo
E. Fermi 2, I-50125 Firenze, Italy}
\altaffiltext{2}{INAF - Osservatorio Astrofisico di Arcetri, Largo
E. Fermi 5, I-50125 Firenze, Italy}
\altaffiltext{3}{Harvard-Smithsonian Center for Astrophysics, 60
Garden Street, Cambridge, MA 02138}
\altaffiltext{4}{INAF - Osservatorio Astronomico di Roma, Via 
Frascati 33, I-00040 Monte Porzio Catone, Italy}
\altaffiltext{5}{Dipartimento di Astronomia, Universit\`a di Padova, 
Vicolo dell'Osservatorio 2, I-35122 Padova, Italy}
\altaffiltext{6}{Astrophysics Science Division, Code 662, NASA/Goddard Space Flight  
Center, Greenbelt, MD 20771, USA}
\altaffiltext{7}{Department of Physics and Astronomy, Johns Hopkins University,  
Baltimore, MD 21218, USA}
\altaffiltext{8}{INAF - Osservatorio Astronomico di Brera, via Brera 28, 
I-20121 Milano, Italy}

\begin{abstract}
We present the results of infrared L-band ($3-4~\mu m$) and M-band ($4-5~\mu m$) VLT-ISAAC spectroscopy of five 
bright Ultraluminous InfraRed Galaxies (ULIRGs) hosting an AGN. From our analysis we distinguish two
 types of sources: ULIRGs where the AGN is unobscured (with a flat continuum and no absorption
 features at $3.4~\mu m$ and $4.6~\mu m$), and those with highly obscured AGNs (with a steep, reddened continuum 
and absorption features due to hydrocarbons and CO). Starburst activity is also present in all of the sources as inferred 
from the $3.3~\mu m$ PAH emission line.
A strong 
correlation is found between continuum slope and CO optical depth, which suggests that deep carbon monoxide 
absorption is a common feature of highly obscured ULIRG AGN. Finally we show that the AGN dominates the $3-4~\mu m$ 
emission, even if its contribution to the bolometric luminosity is small.
\end{abstract}


\section{Introduction}

Determining the role and physical properties of Active Galactic Nuclei (AGNs) in Ultraluminous Infrared Galaxies (ULIRGs) is fundamental for many reasons. ULIRGs are the most luminous sources in the local Universe and dominate the bright end of the infrared counts (Sanders \& Mirabel~1996 and references therein); they represent the local counterpart of the high redshift objects discovered by SCUBA (Barger et al.~1998) that dominate the cosmic sub-millimeter background. Furthermore, \textit{Spitzer} completes the millimeter and sub-~mm surveys with its mid-IR selection function and enhances the sensitivity to warm systems, which shows extreme infrared/optical luminosity ratios (Lonsdale, Farrah \& Smith~2006).
Therefore the study of local ULIRGs can unveil the nature of the high redshift sources, which are too faint to be studied in detail with current facilities. 
Finally, ULIRGs may contain luminous heavily obscured AGNs (by gas with a column density $N_H>10^{24}cm^{-2}$ in X-rays), so it is important to determine the contribution of accretion to their infrared emissivity in order to achieve a complete census of the AGN population in the near Universe.\\
A promising new approach to the study of ULIRGs is infrared L-band ($3-4~\mu m$) spectroscopy. 
A comparison between the spectral energy distributions of AGN and starburst (hereafter SB) shows that, with the same 
bolometric luminosity, the L-band emission by an AGN is a factor of $\sim$100 higher 
than that of a SB (Risaliti et al.~2006, hereafter R06).
Therefore, when studying mixed sources, where both an AGN and a SB are present, 
the emission of the AGN
component can be revealed even if the SB is bolometrically dominant and/or the AGN is heavily obscured.
Unfortunately, the high background and low atmospheric transmission in the L-band have made
high quality spectra of ULIRGs difficult to obtain until recently. Imanishi \& Dudley (2000) studied the physics of
 emission and absorption between 3 and 4 $\mu m$ with low resolution spectra of the brightest ULIRGs
 obtained at 4 meter class telescopes.
More recently, R06 and Imanishi et al. (2006) obtained  L-band spectra of ULIRGs with 8-meter class telescopes (VLT and Subaru). 
These high quality spectra have shown the great power of L-band diagnostics in disentangling AGN and SB components 
in ULIRGs. The main results of those studies can be summarized as follows:\\
- a large equivalent width of the $3.3~\mu m$ PAH emission feature ($EW_{3.3}\sim100$ nm) is typical 
of starburst-dominated sources, while the X-ray radiation of an AGN partially or completely destroys 
PAH molecules (Voit 1992).\\
- A strong absorption feature at $3.4~\mu m$ ($\tau_{3.4}>0.2$) due to aliphatic hydrocarbon grains 
is an indicator of an obscured AGN and, indeed, such a deep absorption requires a point source behind a 
screen of dusty gas (this has been verified both theoretically and observationally, Imanishi and Maloney 2003).\\
- A steep inverted continuum (described by a power law $f_\lambda\propto\lambda^\Gamma$, $\Gamma>1$) 
suggests the presence of an obscured, reddened AGN. Again, a high value of $\Gamma$ implies strong 
dust reddening of a compact source.\\
In this paper we present new L-band spectra of the 5 brightest ULIRGs, which are known from independent observation to host an AGN, and extend our study to the M-band ($4-5~\mu m$), using data obtained with 
ISAAC at the Very Large Telescope (VLT) in Paranal, Chile. Our main purposes are: (a) to measure the 
spectral parameters with unprecedented accuracy by determining the continuum over 
a broader spectral region and with higher signal to noise ratio, (b) to analyze the chemical composition and the physical properties of the absorbing/emitting circumnuclear medium. In particular 
we are searching for molecular transitions originating only in warm dense gas, and associated with the 
presence of an AGN, such as the $P$ ($\Delta J=-1$) and $R$ ($\Delta J=+1$) branches of the 1-0 
rotovibrational CO band near $4.65~\mu m$. In this way we can verify if a deep CO hot gas absorption 
is a peculiar feature present only in IRAS 00183-7111 (Spoon et al. 2004), 
or is insted a common signature of heavily obscured AGNs. Even if narrow emission lines are not our main target, in some cases we are able to identify hydrogen 
atomic and molecular features partially blended with residual sky lines (whose subtraction is incomplete in low resolution spectra, van Dishoeck et 
al.~2003). Finally, (c) we wish to determine the relative SB/AGN contribution to both the $3-4~\mu m$ and the total 
IR emission.\\
In Section 2 we present our sample and review the known physical properties of each 
source, which will be compared with our results presented in the following Sections. 
We present the L and M spectra in Section 3, and analyze them from a physical and chemical point of view in Section 4. 
In Section 5 we discuss our results. More specifically, Section 5.1 discusses the relative AGN/SB contribution to the total luminosity, while the physical properties of the dusty emitter/absorber are illustrated in Section 5.2. 
Our conclusion are summarized in Section 6.\\
Throughout this paper we assume $H_0=70, \Omega_M=0.3, \Omega_\Lambda=0.7$.

\section{The sample}

The sources presented here have been drawn from the sample of Genzel et al.~(1998) consisting of the 15 ULIRGs in the IRAS Bright Galaxy Sample (Sanders et al.~2003). 
One of our sources, IRAS 05189-2524 is not included in the original sample 
but fullfills all its selection criteria (Table \ref{tb:car}).
\begin{table*}[!t]
\begin{center}
 \begin{tabular}{lcccccccccc}
 \hline
 source     & z      & $S_{12}^a$ & $S_{25}^a$ & $S_{60}^a$ & $S_{100}^a$ & $\log L_{IR}^b$ & J$^c$ & H$^c$ & K$^c$ & optical class\\
 \hline
 NGC 6240        & 0.0245 &   0.6      & 3.4        & 22.7       & 27.8   & 11.85           & 10.3  & 9.5  & 9.1    & LINER $^1$ \\
 IRAS 05189-2524 & 0.0426 &   0.73     & 3.44       & 13.94      & 11.36  & 12.16           & 14.0  & 12.7  &11.7   & Sy2 $^2$     \\
 IRAS 19254-7245 & 0.0617 &   0.2      & 1.2        & 5.5        & 5.8    & 12.06           & 14.0  & 12.7  & 11.7  & Sy2 $^3$\\
 IRAS 20551-4250 & 0.0428 &   0.3      & 1.9        & 12.8       & 10.0   & 12.04           & 13.4  & 12.2  & 12.2  & H$_{II}$ $^2$\\
 IRAS 23060+0505 & 0.173  &   0.2      & 0.4        & 1.2        & 0.8    & 12.43           & 14.1  & 12.9  & 11.6  & Sy2  $^4$\\
 \hline
 \end{tabular}
 \end{center}
\caption{\footnotesize{photometric data for the observed sources. Notes: $^a$: flux density in the IRAS filters 
in units of Jy. $^b$: Total infrared luminosity, calculated from Eq.(\ref{Esq:1}) and assuming the concordance 
cosmology ($H_0,\;\Omega_M,\;\Omega_\Lambda$)=(70, 0.3, 0.7), Spergel et al. 2003. $^c$: Infrared Magnitudes 
from Duc, Mirabel, and Maza (1997). References for the optical classification are: $^1$ Veilleux et al. 1995, 
$^2$ Veilleux et al. 1999, $^3$ Mirabel et al. 1991, $^4$} Hutchings $\&$ Neff 1988} 
\label{tb:car}
\end{table*}

The luminosity limit used to define a ULIRG is $L_{IR}>10^{12}L_\odot$, where $L_{IR}$ is the total 
8-1000 $\mu m$ luminosity calculated with the IR flux equation of Sanders and Mirabel (1996):
\begin{equation}
F_{IR}=1.8\times10^{-11}(13.48~S_{12}+5.12~S_{25}+2.58~S_{60}+S_{100})
\label{Esq:1}
\end{equation}
here $F_{IR}$ is the total IR flux in $erg\;s^{-1}cm^{-2}$, and $S_{12},\;S_{25},\;S_{60},\;S_{100}$ 
are the flux densities in the IRAS filters in units of Jy.\\
One source, NGC 6240, has a slightly lower total luminosity. However, it is included in our sample, as in 
G98, because its flux density at $60~\mu m$ exceeds 5.4 Jy by a factor of 4 (as shown in 
Table \ref{tb:car}), moreover all of its morphological and physical properties are typical of ULIRGs.\\
Our selection among the ULIRGs in G98 has been driven by the following criteria:\\
- brightness in the infrared (L \& M) bands;\\
- presence of AGNs with different levels of obscuration;\\
- observability from the VLT site. \\
The final subsample of five ULIRGs have been observed with ISAAC at the VLT with two main motivations:\\
1) We wanted to obtain the first M-band spectra for these ULIRGs, thus covering $2~\mu m$ of spectral range when combined with L-band spectra. 
The large spectroscopic interval and the high quality of these data allow us to 
investigate the continuum shape, the physical properties and the chemical composition of the emitting/absorbing medium with unprecedented precision.\\
2) Our sources are the best studied ULIRGs at other wavelengths. However, in some cases the observations 
do not provide complete information, as in the case of NGC 6240, which shows a double AGN only 
in the X-rays. Our aim is to identify both the AGN and the SB components in the IR spectra, and to estimate their relative 
contribution to the $3-4~\mu m$ and total luminosity.

\subsection{Sample properties}
In this paragraph we review the morphological and spectral properties of our sample.

{\bfseries IRAS 05189-2524}\\
IRAS 05189-2524 has a compact structure, typical of a final merging stage.
Only a single nucleus is visible in the optical images, and the only evidence for merging is a faint tidal tail (Veilleux, Kim \& Sanders 2002). According to optical spectroscopic data, 
IRAS 05189-2524 has been classified as a type 2 Seyfert galaxy (Veilleux et al. 1999). In the X-rays the absorbing medium is optically thin with a column density $N_H\sim10^{23}cm^{-2}$. Even if the AGN dominates the X-ray emission, the low X-ray to infrared ratio indicates that the AGN is
not the dominant energy source (Severgnini et al.~2001).

{\bfseries IRAS 19254-7245}\\
The {\em Superantennae} galaxy is an interacting system with a separation of 10~kpc between the two nuclei, 
and prominent tidal tails.
It is optically classified as a Seyfert 2 (Mirabel et al. 1991).
The infrared emission originates from the central region with no significant contribution from the tails. 
Infrared images obtained with the
ISO satellite shows that the dominant contribution to the total luminosity comes from the southern nucleus, 
which also contains most of the
molecular gas present in the system (Charmandaris et al. 2002).
Polarized spectroscopy reveals that the southern nucleus hosts a heavily absorbed AGN which dominates the L-band emission, contributing for $40-50\%$ to $8-1000~\mu m$ luminosity, while the northern one is in a post starburst phase (Berta, et al. 2003).
The X-rays emission arises predominantly from the southern nucleus (Braito et al.~2003). The X-ray continuum 
is flat and a Fe K$\alpha$ line is present, with EW$\sim 1.4$~keV, typical of a reflection-dominated AGN, 
whereas the direct continuum is completely obscured by a column density N$_H>10^{24}$~cm$^{-2}$. Assuming a Compton-thick model, the X-rays intrinsic luminosity is L$_{2-10\;keV}>10^{44}$erg~s$^{-1}$, implying that the AGN 
contribution to the bolometric luminosity is significant. At lower energies (0.5-2 keV) a thermal component 
is observed (k$_B$T$\sim0.9$~keV), probably related to a compact starburst.

{\bfseries IRAS 20551-4250}\\
IRAS 20551-4250 is a merging system in an early stage (Franceschini et al. 2003). 
It is optically classified as a H II region (Veilleux et al. 1999), while in the mid-IR it resembles a SB galaxy 
(Genzel et al.~1998). The X-rays emission is clearly dominated by an obscured AGN, with 
luminosity L$_{2-10\;keV}\sim~7.0\times10^{42}$erg~$s^{-1}$ and column density N$_H\sim8\times10^{23}$~cm$^{-2}$ 
(Franceschini et al. 2003). 
As in the case of IRAS~19254-7245, the relative AGN contribution to the bolometric luminosity is uncertain, 
but probably significant.

{\bfseries IRAS 23060+0505}\\
This galaxy has been optically classified as a Seyfert~2.
In the infrared its spectral energy distribution and the luminosity (L$_{IR}\sim1.3\times10^{46}$erg~s$^{-1}$) 
are typical of quasars (Hill et al.~1987). Optical images show a complex morphology with prominent tidal tails 
(Hutchings \& Neff~1988). Optical and near-infrared spectropolarimetric studies reveal a reflection continuum with an extinction A$_V\sim 2.9$~mag, combined with unpolarized 
emission extincted by A$_V\sim20$~mag (Young, Hough \& Axon~1996). The X-ray data confirm the presence of a 
moderately obscured (N$_H\sim8\times10^{22}$~cm$^{-2}$) AGN with L$_{2-10\;keV}\sim~1.5\times10^{44}$erg~s$^{-1}$ 
(Brandt et al.~1997).

{\bfseries NGC 6240}\\
NGC~6240 is an interacting system in an initial state of merging showing two distinct nuclei with gaseous and 
stellar tails. It is optically classified as a LINER (Veilleux et al.~1995). In the hard X-rays (E$>10$~keV) 
the high observed luminosity is interpreted as direct AGN emission absorbed by a column density 
N$_H\sim2\times10^{24}$cm$^{-2}$ (Vignati et al.~1999); in the 2-10~keV range the presence of the AGN is indicated 
by a Fe $K\alpha$ line with large EW. Recently, observations obtained with \emph{Chandra} revealed that both nuclei host 
an AGN (Komossa et al.~2003). This has been confirmed by our L-band observations (Risaliti et al.~2006B), which will be summarized in the next Section. The mid-IR continuum of NGC 6240 is typical of a starburst, with an indication 
of the presence of an AGN coming from high ionization emission lines, such as [OIV] $\lambda25.9~\mu$m (Lutz et al.~2003). 
A recent study based on observations performed with {\em Spitzer} confirms this result (Armus et al.~2006) 
and estimates an AGN contribution to the bolometric luminosity of 20-24\%. 

\section{Observations and data reduction}
The L and M-band spectra were obtained with the instrument ISAAC at the UT1 unit of the Very Large Telescope (VLT) on Cerro Paranal, Chile. All L-band (except for two sources) and M-band data were obtained during two nights on July 28-30, 2004. L-band spectra of IRAS 19254-7245 and IRAS 20551-4250 were obtained on June 2-4, 2002.\\
Sky conditions were photometric, with a seeing always of the order of, or better than 1 arcsec in the K-band.
The observations were performed with a 90~arcsec long slit, with width of 0.6 or 1.0 arcsec depending on the seeing (see Table~2 for all the relevant observation details).\\
The total exposure time varied from $\sim$1 hour for the brightest sources to 2 hours for the faintest one.\\ 
\begin{table*}[!t]
  \begin{center}
   \begin{tabular}{rcccclcccc}
   \hline
   Source            & Band & T$^a$ & W$^b$ & $\lambda/\Delta\lambda$ $^c$ & Standard Star & Type & Mag$^d$   \\
   \hline
   IRAS 05189-2524   &  L   & 45   & 1.0 & 360 &   Hip 115033  &  B5V    & 4.80     \\
                     &  M   & 50   & 1.0 & 600 &   HR 07221    &  B5IV   & 4.61    \\
   \hline
   NGC 6240          &  L   & 60   & 0.6 & 600 &   HR 05249    & B2IV-V  & 4.57  \\
                     &  M   & 90   & 0.6 & 800 &   HR 05249    & B2IV-V  & 4.53 \\
   \hline
   IRAS 19254-7245/1 &  L   & 60   & 1.0 & 360 &   Hip 077126  & B3V     & 5.92    \\
                  /2 &  L   & 45   & 1.0 & 360 &   Hip 000183  & B4V     & 5.42   \\
                     &  M   & 90   & 0.6 & 800 &   HR 094703   & B4IV    & 4.94    \\
   \hline
   IRAS 20551-4250   &  L   & 60   & 1.0 & 360 &   Hip 076126  & B3V     & 5.92   \\
                     &  M   & 60   & 0.6 & 800 &   Hip 100881  & B4V     & 5.23  \\
   \hline
   IRAS 23060+0505   &  L   & 60   & 1.0 & 360 &   Hip 115033  & B5V     &  4.80 \\
                  /1 &  M   & 30   & 1.0 & 600 &   Hip 115033  & B5V     &  4.82 \\
                  /2 &  M   & 30   & 1.0 & 600 &   HR 07221    & B5IV    &  4.61 \\
                  /3 &  M   & 60   & 1.0 & 600 &   HR 07221    & B5IV    &  4.61 \\
   \hline
   \end{tabular}
  \end{center}
\caption{Observation log for our ISAAC program. IRAS 19254-7245 was observed twice in the L-band, and 
IRAS 23060+0505 three times in the M-band. Notes: $^a$: observing time in 
minutes. $^b$: slit width (arcsec). $^c$: spectral resolution. $^d$: L/M magnitude of the standard star.}
\label{tb:obs}
\end{table*}
In order to avoid saturation due to the high background ($\sim3.9\;\mbox{mag/arcsec}^2$ in the L-band, 
$\sim1.2\;\mbox{mag/arcsec}^2$ in the M-band) the spectra were taken in chopping mode with single exposures 
of 0.56 s. The spectra were then aligned and merged into a single image. We performed a standard data reduction, 
consisting of flat-fielding, background subtraction, and spectrum extraction, using the IRAF\footnote{The IRAF package is distributed by the National Optical Astronomy Observatory, which is operated by the Association of Universities for Research in Astronomy, Inc., under cooperative agreement with the National Science Foundation.} 2.11 package.\\ 
For each object we also acquired the spectrum of a standard star immediately before or after the target 
observation. The star was chosen to have the same airmass as the target, within 0.1, and with L/M magnitude 
between 4 and 6 (Table~2). This allowed us to obtain high quality spectra with 2 minute long observations. Corrections 
for sky absorptions and instrumental response were obtained from the standard stars spectra, divided by their 
intrinsic emission, assumed to be a pure Raleigh-Jeans in the 3-5 $\mu m$ range (all the standards 
were B stars).\\
In order to facilitate the background subtraction, the observations were performed by nodding the source along 
the slit (with a throw of 15 arcsec). However, the instrumental response is not constant along the slit direction. 
We compensate for this effect by performing a separate reduction for each of the three spectra relative to the three 
different positions along the slit. The results before the division by the three standard stars spectra 
(obtained in the same way) were in general rather different one from the other, both in absolute flux and spectral 
shape. However, after the division by the stars, the three spectra of all sources overlapped within the errors.\\
With the aim of obtaining a precise absolute calibration, we took into account aperture effects by analyzing the 
profiles of both the targets and the standard stars along the slit. We assumed a Gaussian profile and we estimated 
the fraction of flux inside the slit by assuming a perfect centering. The relative error is 
$<10\%$ for the L-band, and $<20\%$ for the M-band. The precision of the absolute calibration is confirmed by 
a) observations of the same source performed in two different nights, with two different standard stars, providing 
final spectra consistent within the errors; b) the agreement between our spectra and those of 
Imanishi \& Dudley (2000) for three common sources; and c) the cross calibration between our L-band and M-band spectra, 
which match within 15\%, as expected from our estimates of the absolute calibration errors.\\
For each source we obtained the final spectrum by merging the single spectra relative to the three slit positions, 
and when two or more observations were present, merging the final calibrated spectra. The errors were estimated 
from the Poissonian noise in the sky counts, which are by far the dominant source of noise (for ease of comparison, 
a typical ULIRG has a magnitude $L\sim11-13$, i.e. at least $\sim600$ times fainter than the sky background in the 
same extraction region).
The relative error due to the standard division is negligible when compared to the former.\\
The atmospheric transmission is not constant in the $3-5~\mu m$ range. In several narrow spectral intervals in
the M band, and one in the L band, the transparency is too low to obtain significant data. We excluded these 
intervals from our spectral analysis. In other wavelength ranges, such as between 2.9 and 3.2 $\mu m$, and in 
the M band in general, the atmospheric conditions vary on short timescales. This implies that the sky features 
are not completely removed with the division by the calibration star. In these intervals of low spectral quality, 
we rebinned the spectral channels above the instrument resolution, in order to have statistically significant 
measurements in each spectral point. 
Finally, for the spectral intervals with good atmospheric transmission and high signal-to-noise ratio,
we chose the rebinning in order to have $\Delta\lambda<30$ nm, which 
is a factor of 2 smaller than the spectral resolution. 

Two out of five sources (IRAS 19254-7245 and NGC~6240) are known mergers.
For these sources we chose the slit angle in order to obtain spectra for both nuclei. In the case of NGC~6240 
we were able to obtain high quality spectra of both nuclei even if their emission wings overlap. A detailed 
description of the data reduction and analysis for NGC~6240 spectra are reported in Risaliti et al. 2006B. For 
IRAS 19254-7245 we obtained a high quality spectrum of the brighter nucleus, which accounts for $\sim$95\% of 
the observed infrared emission, and a low S/N one for the fainter one, sufficient to estimate the flux and 
continuum slope, but not for a detailed spectral analysis.
Therefore, here we do not discuss the fainter nucleus in the following analysis.\\
The final L and M-band spectra of our sources are shown in Figure 1.

\section{Data analysis}
We present and briefly discuss the spectrum of each source, disentangling the AGN and SB contributions by means of L-band 
parameters and comparing our results with those reported in Section 2.1. We fitted our spectra with a power-law continuum
and broad Gaussian absorption and emission features, using the CIAO 3.3 software package (Fruscione et al.~2006). Due to the small wavelength coverage, to the low S/N ratio, and to the presence of broad absorptions, we left the relative normalization between the L and M bands free to vary within the uncertainties given above, i.e. $15\%$.\\

In order to estimate the relative contribution of the AGN to the $3-5~\mu m$ ULIRG emission, we use a 
simple model to disentangle the two components. We assume that the intrinsic spectrum of 
the AGN, $f_\lambda$(AGN), is a power law ($f_\lambda\propto\lambda^\Gamma$) with spectral index $\Gamma_{int}=-0.5$, in agreement with the L-band 
spectra of ULIRGs dominated by unabsorbed AGNs (R06), and with L-band spectra of pure type~1 AGNs (Imanishi \& Wada~2004). 
The observed AGN spectrum is obtained from the intrinsic 
spectrum, absorbed by a wavelength-dependent optical depth, $\tau(\lambda)\propto\lambda^{1.75}$ 
(Cardelli et al.~1989). The intrinsic starburst component, $f_\lambda$(SB), is modelled by a continuum with $\Gamma_{int}=-0.5$ 
(as for unobscured AGNs) plus a broad emission feature at $3.3~\mu m$ with $EW_{SB}=110$~nm. 
This is in agreement with the average observed L-band spectra of starburst-dominated ULIRGs (Imanishi \& Dudley~2000, R06)\footnote{We note that our pure starburst spectra
differ from the average spectra of lower-luminosity
normal galaxies (which have $\Gamma\sim-2$, Lu et al. 2003), probably due
to a larger contribution by hot dust in powerful SB.}. The ULIRG emission, $f_\lambda$, is obtained as the combination of the two components:
\begin{equation}
f_\lambda=\alpha f_\lambda\mbox{(AGN)}e^{-\tau(\lambda)}+(1-\alpha)f_\lambda\mbox{(SB)},
\label{fm:dec}
\end{equation}
where $\alpha$ is the fraction of the L-band luminosity due to the AGN. We express the optical 
depth as $\tau(\lambda)=\tau_L\times (\lambda/3.3~\mu m)^{-1.75}$, where $\tau_L$ is the optical depth 
at $3.3~\mu m$. Therefore, the model has two free parameters, $\alpha$ and $\tau_L$. These parameters and 
their errors can be computed from the observed $\Gamma$ and $EW_{3.3}$ in the composite spectrum as follows: 
\begin{equation}
 \tau_L=\frac{EW_{SB}(\Gamma-\Gamma_{int})}{\beta(EW_{SB}-EW_{3.3})}
\end{equation}
and 
\begin{equation}
 \alpha=\frac{EW_{SB}-EW_{3.3}}{EW_{SB}-EW_{3.3}(1-\mbox{e}^{-\tau_L})},
\end{equation}
where $\beta=1.75$ is the slope of extinction curve in the L-band.

An additional source of systematic errors is the uncertainty on the intrinsic AGN and SB spectral 
slope, and, more important, on the intrinsic EW$_{SB}$ in SBs, which is found to span from $\sim80$~nm to $\sim150$~nm in a pure starburst sample (see R06 for further details). If these are taken into account, the true errors could be larger than the final one by 
a factor of $\sim2$ (for further details on this model and on errors estimate, we refer to R06 where the model is also applied to a larger sample of ULIRGs).\\
In order to give a quantitative estimate of this uncertainty, we have listed in Table~4 the model parameters of Eq.~3 and Eq.~4 by assuming different values of the intrinsic PAH equivalent width ($EW_{SB}$). As expected from the Equations, by increasing $EW_{SB}$, $\tau_L$ decreases and $\alpha$ increases. However the parameters are consistent within the errors for all the sources; IRAS~05189-2524 and NGC~6240~S present discrepancies of the order of $3-5~\%$ due to our simple model, which does not reproduce details observed in these sources such as a complex continuum (IRAS~05189-2524), or a broad emission line (the Br$\alpha$ 
in NGC~6240~S). The L-band absorption, $\tau_L$, varies a lot (50\%) only in IRAS~20551-4250, discussed in detail below.

The results of our spectral analysis are summarized in Table~\ref{tb:l}.\\

The above analysis is based on the continuum and emission features. 
The broad absorption features present in our L-band and M-band spectra (Fig.~2) were analyzed separately.

The optical depth of the absorption features were estimated with a Gaussian fit, and correcting for the starburst
continuum dilution, as estimated using the model described above, in order to calculate the optical depth 
with respect to the AGN component only. Specifically, we adopted the following 
estimate:
\begin{equation}
\tau_\lambda = -\ln \frac{(F_\lambda(\mbox{\scriptsize{TOT}})-A_\lambda)-F_\lambda(\mbox{\scriptsize{SB}})}{F_\lambda(\mbox{\scriptsize{AGN}})}=-\ln\frac{F_\lambda(\mbox{\scriptsize{AGN}})-A_\lambda}{F_\lambda(\mbox{\scriptsize{AGN}})}
\end{equation}

where $A_\lambda$ is the Gaussian amplitude, and $F_\lambda(\mbox{\scriptsize{TOT}})$,$F_\lambda(\mbox{\scriptsize{SB}})$ and $F_\lambda(\mbox{\scriptsize{AGN}})$ are
the total, SB and AGN continuum fluxes. All the values refer to the central wavelength of the
absorption feature (3.4~$\mu$m or 4.6~$\mu$m).
 
In order to better investigate the properties of the CO absorption features, we compared our M-band spectra with the 
low 
resolution Spitzer spectrum of IRAS00183-7111 (Spoon et al.~2004) and high resolution ISAAC data of NGC4945 
(Spoon et al.~2003). Due to the low resolution and the high atmospheric noise of the M-band spectra, 
we cannot use a complex analytical model like in the case of IRAS 00183-7111 (Spoon et al.~2004), nor can we try and reproduce the 
M-band features by varying the contributions of  different CO ice and gas components as for NGC~4945 (Spoon et al.~2003 and 
references therein). However, we compared the position 
and the full width at half maximum of the absorption features  with the ones of those high signal-to-noise spectra.
The results for each single source are discussed below.\\

\begin{figure*}[!h]
\begin{center}
\includegraphics[scale=0.40]{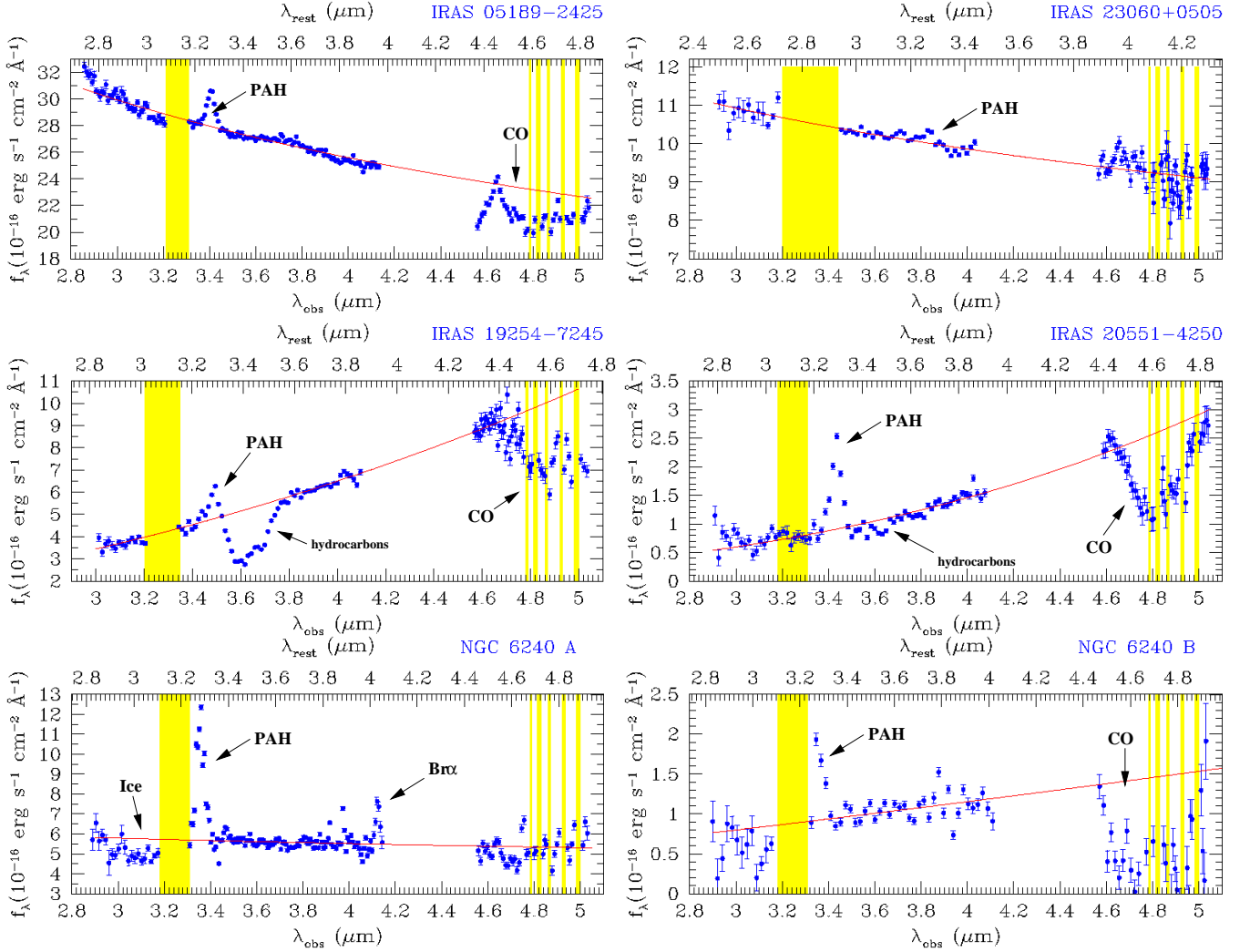}
\end{center}
\caption{$\lambda-f_\lambda$ plot of our sources; spectral points and continuum slope are reported 
and the principal features, with measured physical parameters (see Table~3) are indicated. Bad atmospheric regions are in yellow.}
\label{fg:spettri}
\end{figure*}

{\bfseries IRAS 05189-2524}\\
$\bullet$ The spectral index of the $\lambda-f_\lambda$ continuum is $\Gamma=-0.546\pm0.004$, typical of starburst 
and unobscured AGN (R06). The continuum steepness below $2.9~\mu m$ is interpreted as due to the host galaxy 
emission in the K-band, therefore we have neglected these points in determining the slope. 
The PAH emission at $3.3~\mu m$ has $EW_{3.3}=4.0\pm0.2\;nm$, almost two orders of magnitude smaller than in a 
pure starburst (Imanishi \& Dudley 2000, R06), thus indicating the presence of an AGN. As shown in 
Figure \ref{fg:spettri}, a simple power law appears inadequate to reproduce the details of the continuum emission. 
In particular, the slope $\Gamma$ increases at wavelengths $\lambda > 3.8~\mu m$.
We will not discuss any further these small deviations only visible in the highest S/N cases, and will use 
the average spectral slope in order to analyze in a homogeneous way all the sources in the sample.\\
$\bullet$ 
The M-band spectrum of IRAS~05189-2524 shows possible absorption features due to $\mbox{CO}_2$ (centered at $4.26~\mu m$ rest frame) and CO (Fig.~1 and Fig.~2). 
This interpretation is 
not unique due to the uncertainty in the absolute calibration. It could be possible, within the calibration errors, to 
interpret the spectral points at 4.7-5.0~$\mu$m as the continuum level, and the higher flux points at $\sim4.6~\mu$m as due to an emission feature. 
However, no such a feature is known at the observed wavelengths. Moreover, 
comparing our spectra with literature ones, the positions and FWHM of the two Gaussians reproducing the $4.8~\mu m$ 
absorption profiles are consistent within the errors with Spitzer data (Spoon et al.~2004).
Based on these considerations, 
we interpret the observed spectral shape as due to CO absorption features, and therefore we choose the normalization of the M-band spectrum in order to obtain a good match between the extrapolated L-band 
continuum and the highest flux points in the $4.5-5~\mu m$ interval. With this choice, the cross-calibration factor 
between L-band and M-band is 0.98.
IRAS 05189-2524 shows a spectral profile analogous to IRAS 00183-7111, with an 
optical depth of CO absorption $\tau_{4.6}\sim0.1$. $\mbox{CO}_2$ absorption is also partially observed at the 
lower end of the M-band spectral range. 
Beside broad absorption features, a series of H$_2$ and other atomic emission lines are visible in the M-band 
spectrum. However, it is impossible to estimate their properties for two reasons, both due to the low resolution of our spectra: 
molecular sky lines are not well subtracted, and a forest of fundamental rotovibrational absorption lines of gaseous 
CO (as detected in NGC 4945 by Spoon et al.~2003) are blended with the emission lines.\\
If the measured spectral parameters are interpreted according to the model described above, the 3-5~$\mu m$ emission of 
IRAS 05189-2524 is dominated by an AGN ($\alpha>0.9$) with little or no absorption ($\tau_L<0.03$). A significant 
starburst contribution  is also present, as revealed by the PAH emission line.
On the whole our interpretation is consistent with published data (Imanishi \& Dudley~2000).

{\bfseries IRAS 23060+0505}\\
This is the most luminous and highest redshift source in our sample. The observed spectrum closely resembles that 
of IRAS 05189-2524 (Fig. \ref{fg:spettri}), where the emission of an unobscured AGN dominates the 3-5~$\mu$m 
spectrum.\\
$\bullet$ Despite the high flux of the source, the spectrum is one of the noisiest in our sample, due to bad 
atmospheric conditions during the observation. The only feature present in the L-band spectrum is a weak 
PAH emission with 
$EW_{3.3}=1.8\pm0.7\;nm$. Therefore, an AGN is present and dominates the L-band emission, producing a strong  
dilution of the PAH line.  The continuum spectral slope is 
$\Gamma=-0.36\pm0.02$ consistent with a SB and an unobscured AGN.\\
$\bullet$ No narrow features are detected in the M-band.\\
The normalization factor for the M-band is 1.09, within calibration uncertainties.\\
Applying the analytical model described above we find that the L-band emission is dominated by the AGN emission 
($\alpha>0.9$) with no observed absorption ($\tau_L<0.09$). Our analysis is consistent with the previous 
ones (Section 2.1 and Imanishi \& Dudley 2000).\\

{\bf IRAS 19254-7245}\\
IRAS~19254-7245 is a merging system with two nuclei separated by 10~kpc (Mirabel Lutz \& Maza 1991). In our ISAAC observation the two nuclei are clearly distinguishable. 
However, as reported in Section 2.1, the northern nucleus is too faint to obtain a spectrum with a signal-to-noise ratio good enough for a detailed analysis. Therefore, we will focus on the southern nucleus, accounting for $\sim$95\% of the observed 3-5~$\mu$m emission.
The high quality spectrum of the southern nucleus can be used to analyze the substructures of the absorption features at 3.4~$\mu$m and in the 4.5-5~$\mu$m range. An accurate analysis of the L-band spectrum is presented in Risaliti et al. (2003). Here we report new measurements of the spectral properties obtained by adding the M-band data.\\
$\bullet$ The  L-band spectrum shows both PAH emission, with an equivalent width $EW_{3.3}=24.2\pm1.8\;nm$, 
and two clear absorbed AGN features: a steep continuum ($\Gamma=2.22\pm0.03$), and a strong aliphatic hydrocarbons absorption 
($\tau_{3.4}=0.82\pm0.05$), three times higher than the maximum value expected if the dust is spatially mixed with the 
emitting source ($\tau\sim0.2$, Imanishi \& Maloney 2003).\\
$\bullet$ As in the case of IRAS 05189-2524, the observed $4-5~\mu m$ absorption is ascribed to vibrational 
modes of the CO molecule. We assume that the high flux points at rest frame wavelength $\lambda\sim4.4 \mu$m are at 
the continuum level, and therefore lay on the extrapolation of the L-band continuum (Fig.~1 and Fig.~2). With this choice the 
normalization factor
between the L-band and the M-band is 1.13.\\
Our low resolution spectrum does not allow a detailed analysis of the single substructures. However, it is 
possible to compare the overall absorption profile with that due to gaseous CO 
(such as in IRAS~00183-7111, Spoon et al.~2004) and that due to solid CO (as NGC~4945, 
Spoon et al.~2003). This comparison is further 
complicated by the redshift of our sources, which moves the end of the absorption 
profile out of the M-band (Fig.~1). Despite these limitations, we can qualitatively assess that the 
absorption profile seems to be due to gaseous CO with $\tau_{4.6}=0.44\pm0.04$ (Fig.~2), and we reproduce the absorption profile with two Gaussians 
centered within the errors at the expected position for the P and R-branches of the vibration mode of CO gas.

\begin{figure*}[!h]
\begin{center}
\includegraphics[scale=0.45]{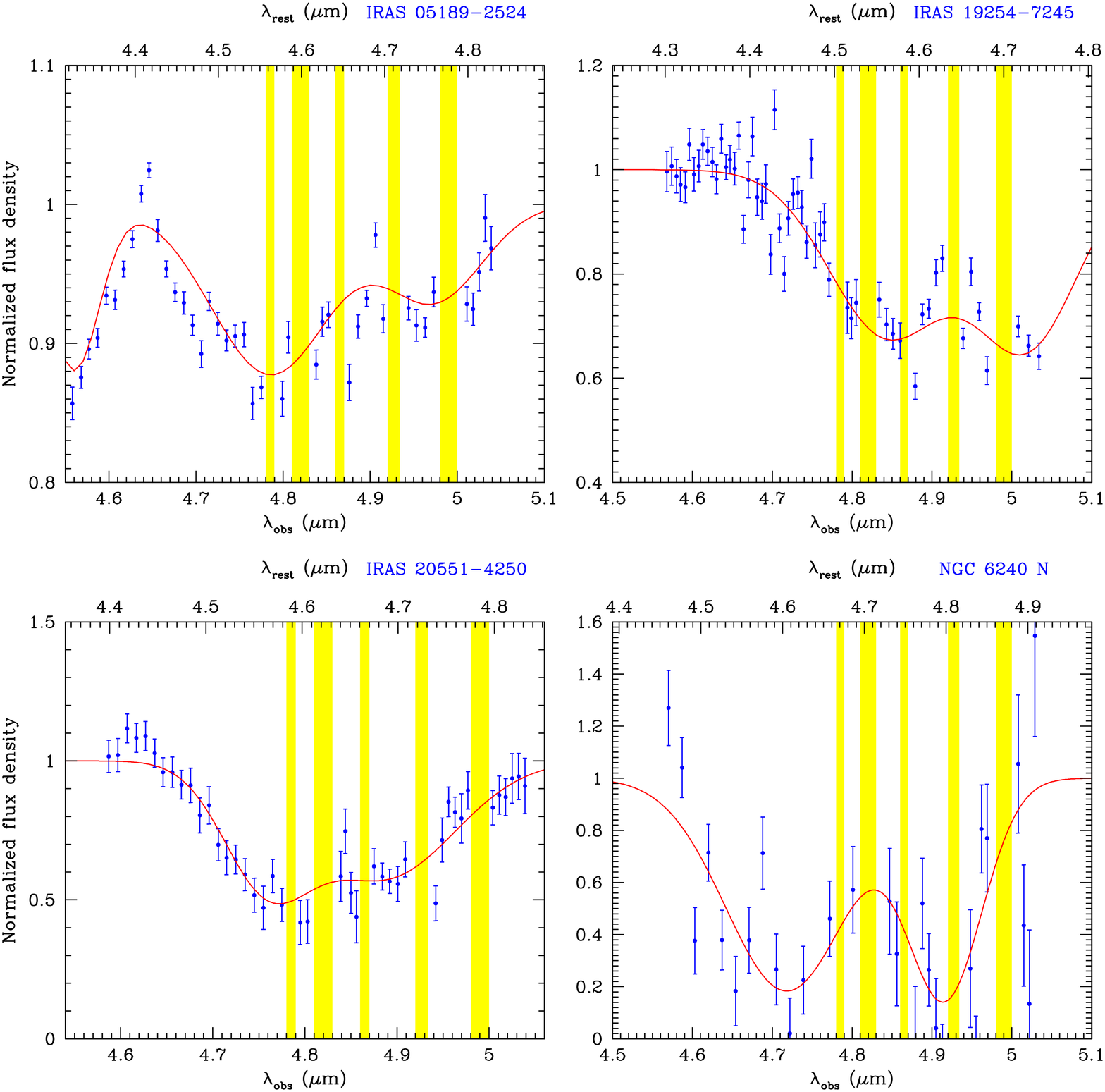}
\end{center}
\caption{Absorption features in the spectra of IRAS~05189-2524, IRAS~19254-7245, IRAS~20551-4250 and NGC~6240: the normalized flux profiles are shown in red. The broad double-branched features centered at $4.67~\mu$m (rest frame) are identified 
with the P- and R- branches of the fundamental vibration mode of CO gas; the red wing of CO$_2$ feature ($\lambda_{rest}=4.26~\mu$m) is detected only in IRAS~05189-2524. 
We obtain these profiles by fitting the M-band data with the proper continuum powerlaw and a set of Gaussian curves.}
\label{fg:assorbimenti}
\end{figure*}

This is also suggested by the lack of H$_2$ emission lines $\lambda4.69~\mu$m, indicating a 
temperature higher than $\sim$2000~K.\\
The analytical model confirms these results: an AGN dominates the L-band emission ($\alpha>0.9$), and it is 
absorbed by a large amount of dust which reddens the spectrum 
with an absorption of $\tau_L\sim2$. This source has been observed in the $3-5~\mu m$ range by our group for the 
first time, and the results are consistent with X-rays observations,which also indicates the presence of a powerful, obscured 
AGN (Braito  et al.~2003).

{\bfseries IRAS 20551-4250}\\
The spectrum of IRAS 20551-4250 (Fig.~\ref{fg:spettri}) is similar to that of IRAS 19254-7345, with a steep inverted 
continuum and deep absorption features. The L-band spectrum has been discussed in R06. Here we add the M-band 
spectrum to the analysis.\\
$\bullet$ The L-band spectrum shows both a $3.3~\mu m$ PAH emission with a large equivalent width $EW_{3.3}=90\pm7\;nm$,
pointing to a powerful SB, and two clear absorbed AGN features: a steep inverted continuum with $\Gamma=3.10\pm0.03$, 
and a significant aliphatic hydrocarbons absorption $\tau_{3.4}=1.5\pm1.1$.
We note that this is the case with the largest correction of $\tau_{3.4}$ due to SB dilution with respect to the observed value
(observed $\tau_{3.4}=0.3\pm0.1$, R06). Also, the wavelength of the absorption peak is shifted to $\sim3.47~\mu$m. This is
probably due to a high abundance of electronegative groups such as -OH and -N0$_2$ in the aliphatic hydrocarbon dust molecules (see R06 for a detailed discussion).
\\
$\bullet$ The M-band features consist of a deep CO absorption and a Pf$\beta$ ($4.65~\mu m$) emission line. Thanks 
to the low redshift ($z=0.0428$) we can observe the P- and R-branches of gaseous CO, in which the absorptions forest 
identified in NGC4945 can not be resolved, however, because of our low spectral resolution combined with the high atmospheric 
variability (Fig.~1 and Fig.~2).
If the whole absorption feature is modeled with one Gaussian broad line, the peak position and FWHM are consistent with gaseous CO and we estimate an optical depth 
$\tau_{4.6}=2.2\pm0.6$. 
Atmospheric turbulence and CO absorption bands prevent us from measuring the profile and equivalent width of the 
Pf$\beta$ line.
The cross correlation factor between L and M bands is 0.83.\\
Our decomposition model confirms that the L-band  spectrum is dominated by an AGN ($\alpha>0.9$) with the 
highest dust reddening ($\tau_L\sim11$) in our sample. 
This high value of $\tau$ is necessary in order to reconcile two apparently contradictory observational results: the high
equivalent width of the 3.3~$\mu$m feature, implying a dominance of the starburst at $\lambda_{rest}=3.3$~$\mu$m, 
and the steep continuum, implying a dominant contribution of an absorbed AGN at $\sim4 \mu$m. 
Anyway, we note that the estimated L-band optical depth is affected by a large relative error, of the order of $40\%$. Furthermore, we note that the estimate of $\tau$ can be strongly affected by our assumption of a fixed value of 
EW$_{3.3}$ with respect to the pure starburst continuum. If we allow for larger values of this parameter we 
increase the fraction of continuum due to the AGN. As a consequence, the observed continuum slope can be 
reproduced with a less absorbed AGN component. For example, if we assume an intrinsic EW$_{3.3}=130$~nm  instead 
of the standard value
of 110~nm (well within the spread observed in pure starbursts, R06) we obtain $\tau=6.7\pm1.3$; varying $EW_{3.3}$ in the range $110-150$~nm we obtain a $\tau_L$ ranging from 11 to 5. 

{\bfseries NGC 6240}\\
VLT data pertaining to NGC~6240 have been discussed in a separate paper (Risaliti et al.~2006B); in the following 
we report a summary of our result for completeness and comparison with the other four bright ULIRGs.
NGC 6240 is the only source in our sample in which we are able to study both nuclei in detail.
Our results confirm the early detection of the double AGN obtained in the hard X-rays with Chandra (Komossa et
al. 2003): in the brightest nucleus (see Figure \ref{fg:spettri}) a broad Br$\alpha$ emission line is detected 
(with a FWHM corresponding to a velocity of $1800\pm200$~km~s$^{-1}$), the only broad line region evidence 
identified in the spectrum of this source. 
The continuum is flat ($\Gamma=-0.17\pm0.03$) and a PAH emission is clearly detected, with $EW_{3.3}=48\pm2$ nm, 
a factor $\sim2$ smaller than in a pure starburst.\\
The faintest nucleus (see Figure~\ref{fg:spettri}, and Figure~2) shows a steep $3-5~\mu m$ emission ($\Gamma=1.28\pm0.17$), 
typical of a reddened AGN (NGC~6240N $\tau_L\sim2$), as in the cases of IRAS~20551-4250 and IRAS~19254-7245, with a strong 
CO absorption feature in the M band ($\tau_{4.6}=1.5\pm1.4$). The only difference between NGC~6240N and the above IRAS sources consists in a small hydrocarbon absorption feature 
(with an upper limit for the optical depth $\sim0.05$),which indicates a possible different dust composition (see discussion in Section 5.2).\\

Summarizing our results (see Table~3), we can interpret the data as follows: the starburst process is always present and an AGN 
is clearly detected
 in all sources. A different amount of dust along our line of sight generates the observed spectral differences.
 In particular the large amount of gas and dust producing the $3.4~\mu m$ and $4.6~\mu m$ absorption bands together 
with a reddened continuum can be related only with a point source behind a screen of dusty gas, such as an AGN. Our M-band spectra show that 
the presence of deep CO absorption features in the 4.6-5~$\mu$m range is always associated with a reddened AGN continuum. 

\begin{table*}
\begin{center}
 \begin{footnotesize}
      \begin{tabular}{lcccccccc}
      \hline
      Source     & $\Gamma$         & $EW_{3.3}$   & $\tau_{3.1}$  & $\tau_{3.4}$  & $\tau_{4.6}$ & $\tau_L$ & $\alpha$ (\%) & $\alpha_{\tiny{BOL}}$ (\%)\\
      \hline
      NGC 6240 S  & $-0.17\pm0.03$   & $48\pm2$    & $0.2\pm0.05$  &   --          & --              & $0.34\pm0.03$   & $64\pm4$ & $<1$  \\
      NGC 6240 N  &  $1.28\pm0.17$   & $50\pm12$   &   --          & $<0.05$       & $1.5\pm1.4$     & $1.4\pm0.1$   & $80\pm20$ & $2\pm1$\\
      I 05189-2524 & $-0.55\pm0.01$  & $4.0\pm0.2$ &   --          & --            & $0.118\pm0.007$ & $<0.03$       & $>95$ & $11.3\pm0.5$ \\
      I 19254-7245 & $2.21\pm0.03$   & $24\pm2$    & --            & $0.82\pm0.05$ & $0.44\pm0.04$   & $1.82\pm0.02$ & $>95$ & $10\pm1$\\
      I 20551-4250 & $3.10\pm0.03$   & $90\pm7$    &  --           & $1.5\pm1.1$   & $2.2\pm0.6$     & $5-11^a$      & $>95$ & $>90$ \\
      I 23060+0505 & $-0.36\pm0.02$  & $1.8\pm0.7$ &   --          &   --          & --              & $<0.09$       & $>95$ & $25\pm8$\\
      \hline
      \end{tabular}
       \end{footnotesize}
      \end{center}
\caption{{\footnotesize Columns 2-5: best fit estimates for the continuum slope 
$\Gamma$ ($f_\lambda\propto\lambda^\Gamma$), the
equivalent width of the 3.3~$\mu$m PAH feature, $EW_{3.3}$ in units of nm, and the optical depths of the absorption features
due to water ice ($\lambda\sim3.1~\mu$m), aliphatic hydrocarbons ($\lambda\sim3.4~\mu$m) and CO ($\lambda\sim3.4~\mu$m). The optical depth values are toward the AGN emission, after subtracting the starburst contribution.
Columns~7-9: estimated values of the continuum extinction $\tau_L$, the contribution of the AGN component to the L-band luminosity ($\alpha$) and to the bolometric luminosity
($\alpha_{\tiny{BOL}}$), obtained, respectively, from Eq.~3, Eq~4 and Eq.~5.}}
\label{tb:l}
\end{table*}
\begin{table*}
\begin{center}
      \begin{tabular}{l|cc|cc|cc}
      & \multicolumn{2}{c}{EW$_{SB}=80$~nm} & \multicolumn{2}{c}{EW$_{SB}=110$~nm} & \multicolumn{2}{c}{EW$_{SB}=150$~nm}  \\
      \hline
      Source          & $\tau_L$  & $\alpha$ & $\tau_L$      &$\alpha$ & $\tau_L$    &  $\alpha$  \\
      \hline                                                                
      NGC 6240 S      & $0.5\pm0.1$ & $51\pm4$   & $0.34\pm0.03$ &$64\pm4$ & $0.3\pm0.03$  &  $74\pm4$    \\
      NGC 6240 N      & $2.7\pm1.3$ & $>90$      & $1.4\pm0.1$   &$80\pm20$    & $1.5\pm0.3$   &  $>90$       \\
      IRAS 05189-2524 & $<0.03$     &  $>95$     & $<0.03$       &$>95$    &  $<0.01$      &  $>95$       \\
      IRAS 19254-7245 & $2.2\pm0.1$ & $>95$      & $1.82\pm0.02$ &$>95$    & $1.85\pm0.05$ &  $>95$       \\
      IRAS 20551-4250 & -           & -          & $8\pm3$       &$>95$    & $5.1\pm0.7$   &  $>95$       \\
      IRAS 23060+0505 & $<0.08$     & $>95$      & $<0.09$       &$>95$    & $<0.08$       &  $>95$       \\
      \hline
      \end{tabular}
      \end{center}
\caption{{\footnotesize Comparison between the best fit  values of $\alpha$ and $\tau$ obtained assuming three different values of $EW_{SB}$
(see text for details).
}}
\label{tb:conf}
\end{table*}

\section{Discussion}
The analysis of our 3-5~$\mu$m spectra leads to the following important results: \\
$\bullet$ Assuming fixed AGN and starburst intrinsic templates, and allowing for a free relative contribution of the two components and a free absorption of the AGN component we can evaluate quantitatively the AGN and starburst contribution to the $3-5~\mu$m flux in all sources.\\
$\bullet$ In all sources with clear indication of an obscured AGN in the continuum spectrum, we find strong gaseous CO absorption features in the M-band. 

\subsection{Estimates of the AGN and SB contributions}

The estimates of the relative contributions of the AGN and starburst components to the L-band flux, presented in 
Section~4 and listed in Table~3, can be complemented with analogous estimates for the contributions to the
total luminosity.

In order to achieve this, we need to take into account
 the L-band to bolometric ratios of pure starbursts, $R_{SB}$, and pure AGNs, $R_{AGN}$.
From the analysis of a large sample of nearby ULIRGs (R06,  
 Risaliti, Imanishi \& Sani submitted) we estimate $R_{SB}\sim 5\times10^{-3}$ and $R_{AGN}\sim1$. The uncertainties
on these values, estimated by the spread of the single objects with respect to the average ratios, are of the
order of 30\%. 
While there are no other ways to determine $R_{SB}$, it is possible to compare our estimate of $R_{AGN}$ with
the value obtained from quasar SEDs (e.g. Elvis et al.~1994). From these average SEDs, based on local PG quasars,
we obtain $R_{AGN}\sim0.5$\footnote{Assuming that in ULIRGs all the intrinsic emission is
reprocessed in the IR, we estimated the bolometric luminosity in QSOs from their total intrinsic emission, from near-IR to X-rays.}, which is consistent with our estimate, considering (a) the large spread in 
quasar SEDs (Elvis et al~1994) and that recent works based on Spitzer data show that the SED of PG quasars
is also heavily contaminated by a starburst component (Netzer et al.~2007).  

\begin{figure*}[!h]
\begin{center}
\includegraphics[scale=0.70]{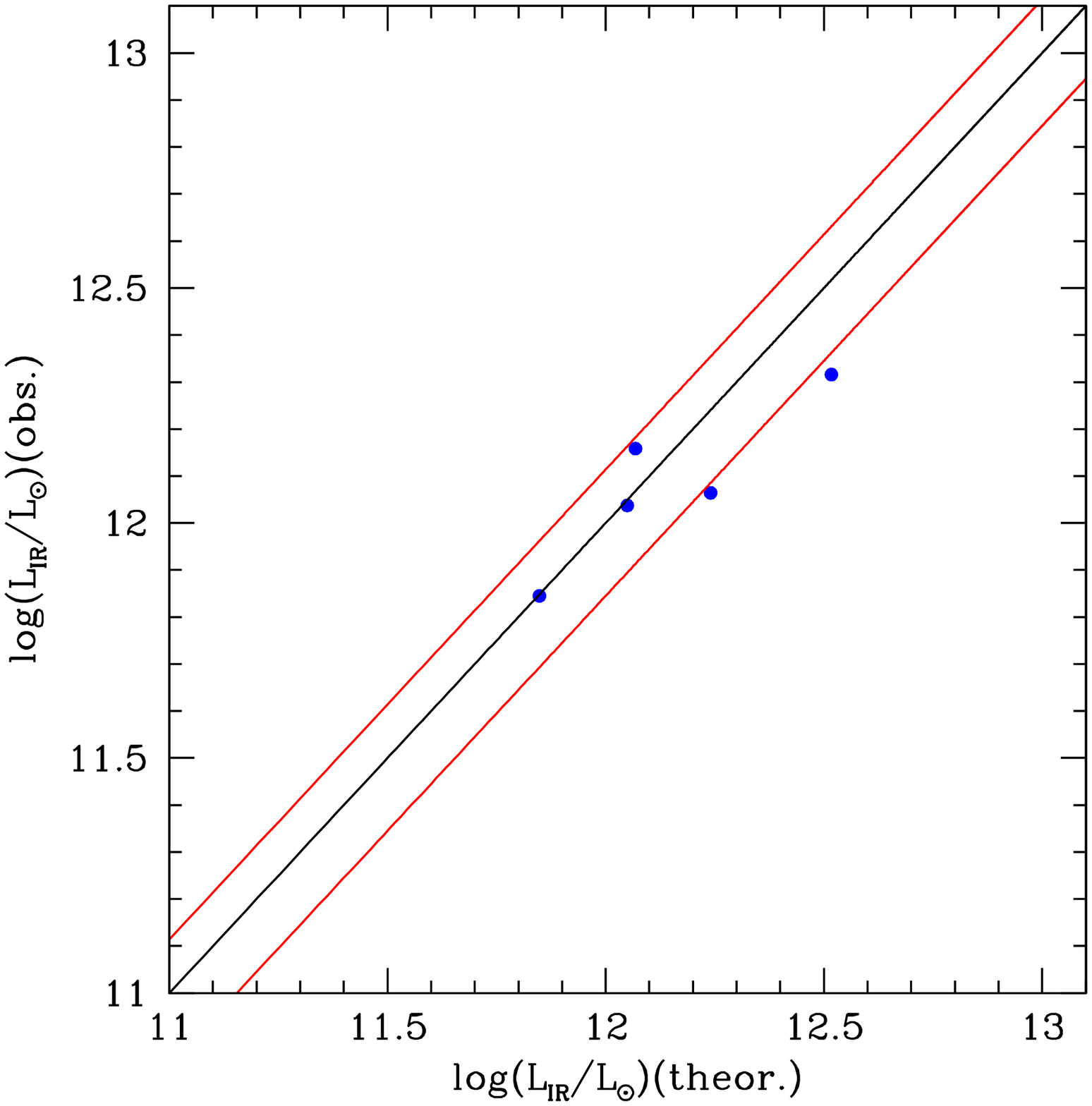}
\end{center}
\caption{Comparison between the total IR luminosity inferred with our model
from the observed L-band emission, assuming bolometric correction (see text for details), and the luminosity measured
by IRAS. Red lines mark the $\pm30\%$ relative error.}
\label{fg:flussi}
\end{figure*}

The correction factor for the AGN contribution to bolometric luminosity is:
\begin{equation}
\alpha_{\mbox{\tiny{BOL}}}=\alpha/(\alpha+K(1-\alpha)).
\label{fm:abol}
\end{equation}
where $K=R_{AGN}/R_{SB}\sim200$.
The errors on this estimate are larger than those on $\alpha$, because of the high uncertainty on K (a factor of $\sim2$). 
However,
it is easily seen from Eq.~\ref{fm:abol} that this does not affect our main conclusion, which 
is $\alpha_{\mbox{\tiny{BOL}}}<1$, provided that K$\gg1$. The results are summarized in Table~\ref{tb:l} and Table~4.\\
To check the consistency of our model, and of Eq.~6, we estimate the expected bolometric flux and compare it 
with the IRAS one. The expected flux is given by:
\begin{equation}
 F_{\mbox{\tiny{BOL}}}=\frac{R_{\mbox{\tiny{AGN}}}\alpha+R_{\mbox{\tiny{SB}}}(1-\alpha)}{\alpha~e^{-\tau}+(1-\alpha)}F(3.3),
\label{eq:exp}
\end{equation}
where $F(3.3)$ is the observed continuum flux at 3.3~$\mu$m. 
Fig.\ref{fg:flussi} show that the two values are in good agreement.

\subsection{Physical properties of the dusty absorber}

The presence of a dusty absorber covering the AGN component is revealed by three different observational
features:\\
$\bullet$ The slope of the AGN component, indicating the reddening of the continuum emission.\\
$\bullet$ The absorption feature at 3.4~$\mu$m due to aliphatic hydrocarbon dust grains.\\
$\bullet$ The absorption features in the 4.5-5.0~$\mu$m range due to CO gas.\\
Considering the L and M bands absorption features in our five sources (Table~3) we note that the presence of deep CO 
absorption ($\tau_{4.6}>0.15$) is always connected with the $3.4~\mu m$ feature (with the exception of the fainter nucleus of NGC 6240, for which we can estimate only the $\tau_{3.4}$ upper limit), consistent with our physical interpretation.
Moreover, these absorption structures are always associated with a steep continuum, which again is a signature of
an obscured AGN. 

Lutz et al. (2004) found no signatures of strong CO absorption both in type 1 and type 2 AGNs, and suggested 
that the deep CO absorption features detected in two ULIRGs (Spoon et al.~2003, 2004) were not directly related to
the obscuring torus responsible for the type~1/type~2 optical classification.  

Here we confirm that such features are signatures of the circumnuclear region of highly obscured ULIRGs. 
An exception is IRAS 05189-2524, which is not heavily absorbed in the continuum but shows a weak 
CO absorption feature 
($\tau_{4.6}<0.1$). However, this optical depth is below the detection limit for most
of the AGN analyzed by Lutz et al.~(2004) with ISO. 
As a consequence, it is possible that CO absorption is common among type~2~AGN, but with an optical depth
significantly smaller than the typical values found in ULIRGs.

A more detailed analysis of the relation between continuum and line/band absorption can be performed by 
checking for possible quantitative correlations between CO and hydrocarbon absorption features and continuum reddening of the AGN component.

\begin{figure*}[!h]
\begin{center}
\includegraphics[scale=0.70]{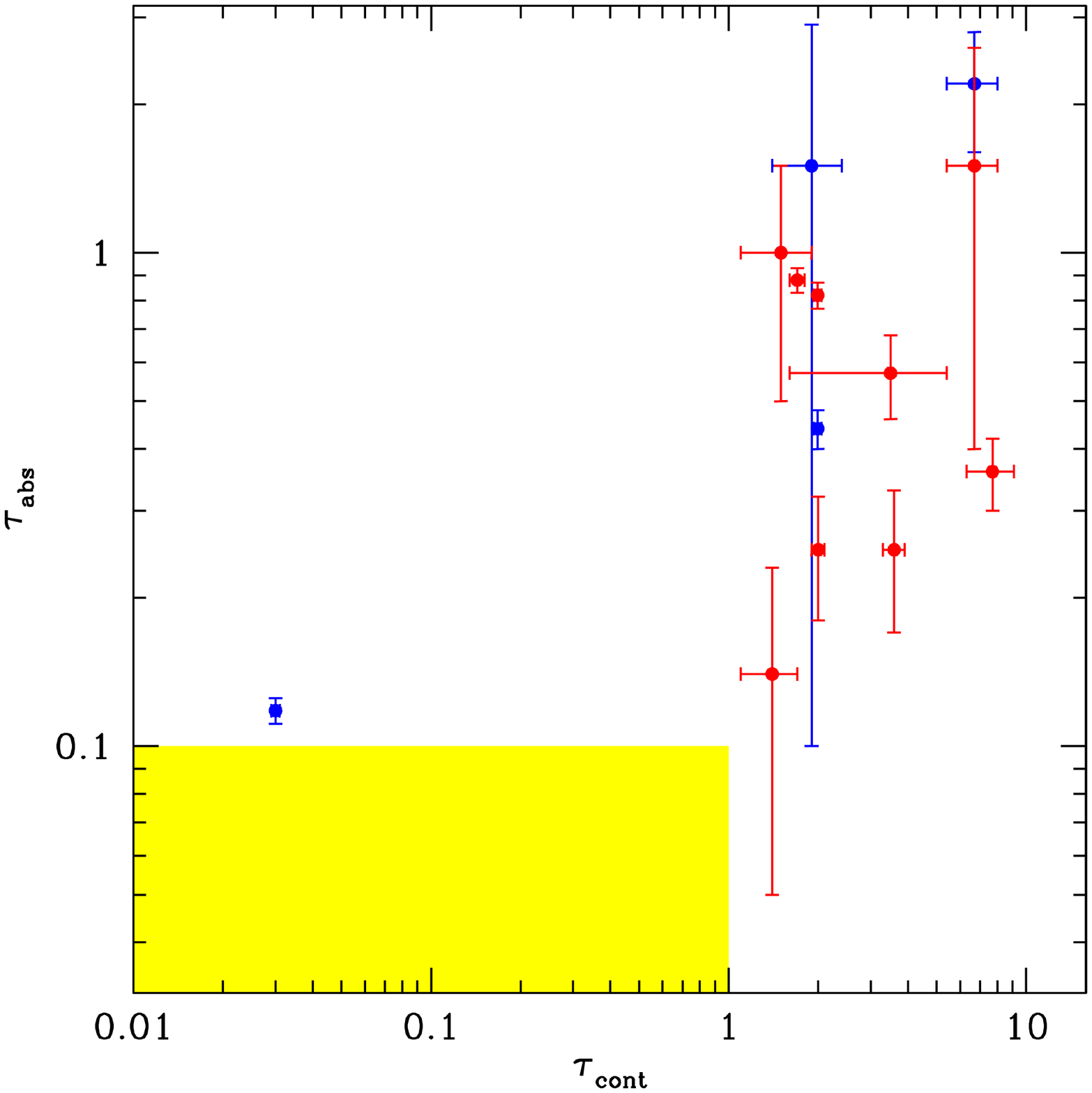}
\end{center}
\caption{Continuum reddening versus absorption features optical depth. The hydrocarbon data are shown in red, in blue are the CO measurements; the isolated blue point corresponds to IRAS~05189-2524,which show a low L-band absorption and a relatively low $\tau_{4.6}$ (see Table~3 and Section~5.2). Spectra with no continuum reddening nor absorption lines lie in the yellow band.}
\label{fg:tau}
\end{figure*}

Figure~4 relates the continuum reddening, $\tau_L$, versus the CO and $3.4~\mu$m hydrocarbon
optical depths, for all the objects with significant
detection of at least one of these localized absorption features. 
The data are from our sample, from R06 and from a re-analysis of the Imanishi et al.~2006 sample using
our model (Risaliti, Imanishi \& Sani submitted).
The figure confirms the correlation discussed above: all the objects with detected absorption features
have reddened continua, while all objects without absorption features are not heavily reddened.
However, we note that no correlation is present within the heavily obscured sources between continuum reddening
and absorption features, nor between the two absorption features themselves.

This suggests that the dust composition is not constant among ULIRGs: a fixed dust composition would
imply a linear correlation between all absorption effects. We plan to investigate the variations in dust
composition complementing our data with Spitzer IRS observations (Sani et al.~2007, in prep.).

\section{Conclusions}
We have presented 3-5 $\mu m$ high quality, low resolution spectra of the five brightest ULIRGs containing an AGN, 
and we have studied the physical properties of the circumnuclear medium which affects the relative AGN/SB observed 
contributions. The high statistic of the data, together with the large wavelength range, allows us to measure spectral diagnostics
($EW_{3.3}$, $\Gamma$, $\tau_{3.4}$ and $\tau_{4.6}$) with unprecedented precision. In our sample we can distinguish 
nuclei with an unobscured AGN (low $EW_{3.3}$, flat continuum and absence of strong absorption) in IRAS 05189-2524, 
IRAS 23060+0505 and the brighter nucleus of NGC 6240, and sources with a heavily absorbed AGN (significant $EW_{3.3}$,
steep inverted continuum with strong dust and gas absorptions) in IRAS 19254-7245, IRAS 20551-4250 and the fainter 
nucleus of NGC 6240. The detection of PAH emission line points to the presence of starburst activity in all sources.\\
To quantify our findings we used a simple analytical model for the deconvolution of the AGN 
and starburst contributions to the bolometric luminosity. All galaxies are
dominated by the AGN component in the L-band,  
while the bolometric emission is dominated by starburst emission in all cases, with the possible exception of IRAS 20551-4250.\\
For the first time we detect strong CO absorption in sources with heavily reddened continua.\\
The lack of correlation between continuum reddening and CO/hydrocarbon absorption features within
heavily obscured sources (see Fig.~4) suggests that
the dust composition is highly variable from source to source.\\

\textit{Acknowledgments}. We are grateful to the anonymous referee for very useful
comments which significantly improved this work. 
This publication makes use of data products from the
2MASS, which is a joint project of the University of Massachusetts
and the Infrared Processing and Analysis Center/California Institute
of Technology, funded by the National Aeronautics and Space Administration
and the National Science Foundation. Also, we made
use of theNASA/IPAC Extragalactic Database (NED) which is operated
by the Jet Propulsion Laboratory, California Institute of Technology,
under contract with the National Aeronautics and Space
Administration.

\appendix
\section{Derivation of $\tau_L$ and $\alpha$}

In the following we derive Eq.~(3) and Eq.~(4) which express the model parameters $\alpha$ and $\tau_L$ as a function of the observed EW$_{3.3}$ and $\Gamma$. The intrinsic EW$_{SB}$ is defined as:
\begin{equation}
 EW_{SB}=\frac{\int_\lambda f_{PAH}~d\lambda}{(1-\alpha)\lambda^{\Gamma_{int}}},
\end{equation}
while, considering that the AGN continuum dilutes the PAH EW, the observed values is:
\begin{equation}
 EW_{3.3}=\frac{\int_\lambda f_{PAH}~d\lambda}{\alpha\lambda^{\Gamma_{int}}+(1-\alpha)\lambda^{\Gamma_{int}}},
\end{equation}
where $\Gamma_{int}=-0.5$ as shown in Section~4.
From Eq.~(A1) and Eq.~(A2), we obtain:
\begin{equation}
 EW_{3.3}=\frac{EW_{SB}~(1-\alpha)}{\alpha e^{-\tau_L}+(1-\alpha)}.
\end{equation}
The observed spectral index $\Gamma$ is obtained from the continuum between $\lambda_1=3.3~\mu$m and $\lambda_2=4~\mu$m (corresponding to the wavelength range with the lowest background noise):
\begin{equation}
 \left\{
\begin{array}{ll}
\Gamma\simeq\frac{\ln (f_{\lambda_1}/f_{\lambda_2})}{\ln \lambda_1/\lambda_2}  \\
\frac{f_{\lambda_2}}{f_{\lambda_1}}=\left(\frac{\lambda_2}{\lambda_1}\right)^{\Gamma_{int}}\frac{\alpha e^{-\tau(\lambda_2)}+(1-\alpha)}{\alpha e^{-\tau(\lambda_1)}+(1-\alpha)}.
\end{array}
\right.
\end{equation}
A first order 
linearization of the above equations, 
based on the the approximation:\\ 
$\tau(\lambda)=\tau_L\left( \frac{\lambda}{\lambda_L}\right) ^{-\beta}\simeq\tau_L\left( 1-\beta\frac{\lambda_2-\lambda_1}{\lambda_1}\right) $, 
($\beta=1.75$) leads to:
\begin{equation}
 \Gamma=\Gamma_{int}+\frac{\beta~\alpha\tau_L e^{-\tau_L}}{\alpha e^{-\tau_L}+(1-\alpha)}.
\end{equation}
We obtain Eq.~(3) and Eq.~(4) by inverting the system formed by Eq.~(A3) and Eq.~(A5).
A numerical solution of the system without the linear approximation of the extinction term
leads to values very similar (and always consistent within the errors) to those estimated
with Equations 3 and 4.




\begin{thebibliography}{}

\bibitem[Armus et al.(2006)]{2006ApJ...640..204A} Armus, L., et al.\ 2006, \apj, 640, 204 

\bibitem[Barger et al.(1998)]{1998Natur.394..248B} Barger, A.~J., Cowie, L.~L., Sanders, D.~B., Fulton, E., Taniguchi, Y., Sato, Y., Kawara, K., \& Okuda, H.\ 1998, \nat, 394, 248 

\bibitem[Berta et al.(2003)]{2003A&A...403..119B} Berta, S., Fritz, J., Franceschini, A., Bressan, A., \& Pernechele, C.\ 2003, \aap, 403, 119 

\bibitem[Bohlin et al.(1978)]{1978ApJ...224..132B} Bohlin, R.~C., Savage, B.~D., \& Drake, J.~F.\ 1978, \apj, 224, 132 


\bibitem[Braito et al.(2003)]{2003A&A...398..107B} Braito, V., et al.\ 2003, \aap, 398, 107 

\bibitem[Brandt et al.(1997)]{1997MNRAS.290..617B} Brandt, W.~N., Fabian, A.~C., Takahashi, K., Fujimoto, R., Yamashita, A., Inoue, H., \& Ogasaka, Y.\ 1997, \mnras, 290, 617

\bibitem[Cardelli et al.(1989)]{1989ApJ...345..245C} Cardelli, J.~A., Clayton, G.~C., \& Mathis, J.~S.\ 1989, \apj, 345, 245 

\bibitem[Charmandaris et al.(2002)]{2002A&A...391..429C} Charmandaris, V., et al.\ 2002, \aap, 391, 429 

\bibitem[Duc et al.(1997)]{1997A&AS..124..533D} Duc, P.-A., Mirabel, I.~F., \& Maza, J.\ 1997, \aaps, 124, 533 

\bibitem[Elvis et al.(1994)]{1994ApJS...95....1E} Elvis, M., et al.\ 1994, \apjs, 95, 1

\bibitem[Franceschini et al.(2003)]{2003MNRAS.343.1181F} Franceschini, A., et al.\ 2003, \mnras, 343, 1181

\bibitem[Fruscione et al.(2006)]{2006SPIE.6270E..60F} Fruscione, A., et 
al.\ 2006, \procspie, 6270

\bibitem[Genzel et al.(1998)]{1998ApJ...498..579G} Genzel, R., et al.\ 1998, \apj, 498, 579 

\bibitem[Hill et al.(1987)]{1987ApJ...316L..11H} Hill, G.~J., Wynn-Williams, C.~G., \& Becklin, E.~E.\ 1987, \apjl, 316, L11 

\bibitem[Hutchings \& Neff(1988)]{1988AJ.....96.1575H} Hutchings, J.~B., \& Neff, S.~G.\ 1988, \aj, 96, 1575 

\bibitem[Imanishi et al.(2006)]{2006ApJ...637..114I} Imanishi, M., Dudley, C.~C., \& Maloney, P.~R.\ 2006, \apj, 637, 114 

\bibitem[Imanishi \& Dudley(2000)]{2000ApJ...545..701I} Imanishi, M., \& Dudley, C.~C.\ 2000, \apj, 545, 701 

\bibitem[Imanishi \& Maloney(2003)]{2003ApJ...588..165I} Imanishi, M., \& Maloney, P.~R.\ 2003, \apj, 588, 165 

\bibitem[Imanishi \& Wada(2004)]{2004ApJ...617..214I} Imanishi, M., \& 
Wada, K.\ 2004, \apj, 617, 214 

\bibitem[Komossa et al.(2003)]{2003ApJ...582L..15K} Komossa, S., Burwitz, V., Hasinger, G., Predehl, P., Kaastra, J.~S., \& Ikebe, Y.\ 2003, \apjl, 582, L15 

\bibitem[Lonsdale et al.(2006)]{2006astro.ph..3031L} Lonsdale, C., Farrah, 
D., \& Smith, H.\ 2006, ArXiv Astrophysics e-prints, arXiv:astro-ph/0603031 

\bibitem[Lu et al.(2003)]{2003ApJ...588..199L} Lu, N., et al.\ 2003, \apj, 
588, 199 

\bibitem[Lutz et al.(2004)]{2004A&A...426L...5L} Lutz, D., Sturm, E., Genzel, R., Spoon, H.~W.~W., \& Stacey, G.~J.\ 2004, \aap, 426, L5 

\bibitem[Lutz et al.(2003)]{2003A&A...409..867L} Lutz, D., Sturm, E., Genzel, R., Spoon, H.~W.~W., Moorwood, A.~F.~M., Netzer, H., \& Sternberg, A.\ 2003, \aap, 409, 867 

\bibitem[Lutz et al.(2002)]{2002A&A...396..439L} Lutz, D., Maiolino, R., Moorwood, A.~F.~M., Netzer, H., Wagner, S.~J., Sturm, E., \& Genzel, R.\ 2002, \aap, 396, 439 

\bibitem[Lutz et al.(2000)]{2000ApJ...530..733L} Lutz, D., et al.\ 2000, \apj, 530, 733 

\bibitem[Maccacaro et al.(1982)]{1982ApJ...257...47M} Maccacaro, T., Perola, G.~C., \& Elvis, M.\ 1982, \apj, 257, 47 

\bibitem[Maiolino et al.(2001)]{2001A&A...365...37M} Maiolino, R., Marconi, A., \& Oliva, E.\ 2001, \aap, 365, 37 

\bibitem[Maiolino et al.(2001)]{2001A&A...365...28M} Maiolino, R., Marconi, A., Salvati, M., Risaliti, G., Severgnini, P., Oliva, E., La Franca, F., \& Vanzi, L.\ 2001, \aap, 365, 28 

\bibitem[Mirabel et al.(1991)]{1991A&A...243..367M} Mirabel, I.~F., Lutz, D., \& Maza, J.\ 1991, \aap, 243, 367 

\bibitem[Pendleton et al.(1994)]{1994ApJ...437..683P} Pendleton, Y.~J., Sandford, S.~A., Allamandola, L.~J., Tielens, A.~G.~G.~M., \& Sellgren, K.\ 1994, \apj, 437, 683 

\bibitem[Ptak et al.(2003)]{2003ApJ...592..782P} Ptak, A., Heckman, T., Levenson, N.~A., Weaver, K., \& Strickland, D.\ 2003, \apj, 592, 782 

\bibitem[Risaliti et al.(2000)]{2000A&A...356...33R} Risaliti, G., Maiolino, R., \& Bassani, L.\ 2000, \aap, 356, 33 

\bibitem[Risaliti et al.(2003)]{2003ApJ...595L..17R} Risaliti, G., et al.\ 2003, \apjl, 595, L17 

\bibitem[Risaliti \& Elvis(2005)]{2005ApJ...629L..17R} Risaliti, G., \& Elvis, M.\ 2005, \apjl, 629, L17

\bibitem[Risaliti et al.(2006)]{2006MNRAS.365..303R} Risaliti, G., et al.\ 2006, \mnras, 365, 303 

\bibitem[Risaliti et al.(2006)]{2006ApJ...637L..17R} Risaliti, G., et al.\ 2006B, \apjl, 637, L17 
 
\bibitem[Sanders \& Mirabel(1996)]{1996ARA&A..34..749S} Sanders, D.~B., \& Mirabel, I.~F.\ 1996, \araa, 34, 749 

\bibitem[Sanders et al.(2003)]{2003AJ....126.1607S} Sanders, D.~B., Mazzarella, J.~M., Kim, D.-C., Surace, J.~A., \& Soifer, B.~T.\ 2003, \aj, 126, 1607 

\bibitem[Severgnini et al.(2001)]{2001A&A...368...44S} Severgnini, P., Risaliti, G., Marconi, A., Maiolino, R., \& Salvati, M.\ 2001, \aap, 368, 44 

\bibitem[Spergel et al.(2003)]{2003ApJS..148..175S} Spergel, D.~N., et al.\ 2003, \apjs, 148, 175 

\bibitem[Spoon et al.(2003)]{2003A&A...402..499S} Spoon, H.~W.~W., Moorwood, A.~F.~M., Pontoppidan, K.~M., Cami, J., Kregel, M., Lutz, D., \& Tielens, A.~G.~G.~M.\ 2003, \aap, 402, 499 

\bibitem[Spoon et al.(2004)]{2004ApJS..154..184S} Spoon, H.~W.~W., et al.\ 2004, \apjs, 154, 184 

\bibitem[van Dishoeck et al.(2003)]{2003Msngr.113...49V} van Dishoeck, E.~F., et al.\ 2003, The Messenger, 113, 49 

\bibitem[Veilleux et al.(2002)]{2002ApJS..143..315V} Veilleux, S., Kim, D.-C., \& Sanders, D.~B.\ 2002, \apjs, 143, 315 

\bibitem[Veilleux et al.(1999)]{1999ApJ...522..113V} Veilleux, S., Kim, D.-C., \& Sanders, D.~B.\ 1999, \apj, 522, 113 

\bibitem[Veilleux et al.(1995)]{1995ApJS...98..171V} Veilleux, S., Kim, D.-C., Sanders, D.~B., Mazzarella, J.~M., \& Soifer, B.~T.\ 1995, \apjs, 98, 171 

\bibitem[Vignati et al.(1999)]{1999A&A...349L..57V} Vignati, P., et al.\ 1999, \aap, 349, L57 

\bibitem[Young et al.(1996)]{1996MNRAS.280..291Y} Young, S., Hough, J.~H., Axon, D.~J., Ward, M.~J., \& Bailey, J.~A.\ 1996, \mnras, 280, 291

\bibitem[Voit(1992)]{1992MNRAS.258..841V} Voit, G.~M.\ 1992, \mnras, 258, 
841 
\end{thebibliography}
\end{document}